# Non-Perturbative Renormalization of the Lattice Heavy Quark Classical Velocity

Jeffrey E. Mandula[a] and Michael C. Ogilvie[b]

[a]Department of Energy, Division of High Energy Physics
Washington, DC 20585

[b]Department of Physics, Washington University
St. Louis, MO 63130

We discuss the renormalization of the lattice formulation of the Heavy Quark Effective Theory (LHQET). In addition to wave function and composite operator renormalizations, on the lattice the classical velocity is also renormalized. The origin of this renormalization is the reduction of Lorentz (or O(4)) invariance to (hyper)cubic invariance. We present results of a new, direct lattice simulation of this finite renormalization, and compare the results to the perturbative (one loop) result. The simulation results are obtained with the use of a variationally optimized heavy-light meson operator, using an ensemble of lattices provided by the Fermilab ACP-MAPS collaboration.

## 1. INTRODUCTION

In the lattice formulation of the Heavy Quark Effective Theory (LHQET), a new renormalization beyond those of the continuum theory arises. It is the the renormalization of the "classical velocity". In the continuum, the classical velocity $v$ is a fixed parameter which appears in the decomposition of the momentum of a heavy particle and the reduced Dirac equation of the heavy quark field.

$$P = Mv + p$$
$$-iv \cdot D\, h^{(v)}(x) = 0 \qquad (1)$$

In the continuum, the second relation is derived from the first, and the velocity that appears in these two contexts is the same. However, on the lattice this is not the case. As we have explained elsewhere, the origin of the difference between the values of $v$ in these two contexts is the reduction of continuum Lorentz (or Euclidean O(4)) invariance to hypercubic invariance on the lattice.[1]

The evaluation of this renormalization is necessary in order to relate calculations performed in the LHQET to physical quantities. One of the most basic of such quantities is the slope of the Isgur-Wise form factor, $\xi(v \cdot v')$ at the origin. In the $M_Q \to \infty$ limit, $\xi$ describes all semi-leptonic decays of heavy mesons into one another, such as the process $B \to D^* l \nu$. The renormalization of $v$ is directly reflected in the value of this slope.

In this talk, we will describe a non-perturbative calculation of the leading multiplicative renormalization of $v$ by means of a lattice simulation. We will first outline how the renormalized value of the classical velocity can be extracted from a lattice simulation, and then describe a variational procedure for enhancing the quality of the signal one obtains in the simulation. Finally, we will give the results, and compare them to the results of from a one-loop perturbative calculation. It is useful to develop the calculation in a power series in the "bare" classical velocity, and we follow this procedure.

## 2. THE PHYSICAL CLASSICAL VELOCITY

The heavy quark effective theory is efficiently formulated[2] by factoring the $M \to$ singular behavior from the field of a heavy quark, leaving a reduced operator

$$h^{(v)}(x) = e^{-iMv \cdot x} \frac{1 + \gamma \cdot v}{2} \psi(x) \qquad (2)$$

There is an independent reduced field for each value of the classical velocity. To define the classical velocity non-perturbatively without global gauge fixing, it is useful to consider a gauge invariant quantity containing the reduced heavy quark field, for example a composite operator for a heavy meson made of one light and one heavy quark. The asymptotic behavior of its propagator will be

$$\Delta^{(v)}(t,\vec{p}) \sim C^{(v)}(\vec{p}) e^{-E^{(v)}(\vec{p})t} \qquad (3)$$

where its rate of fall-off, relative to the $\vec{p} = 0$, $\tilde{v} = 0$ heavy quark energy, is given by

$$E^{(v)}(\vec{p}) = \lim_{M \to \infty} \sqrt{(M+m)^2 + ((M+m)\vec{v}^{(phys)} + \vec{p})^2} - M v_0^{(phys)} \qquad (4)$$

$$= m v_0^{(phys)} + \tilde{v}^{(phys)} \cdot \vec{p}$$

Its first derivative with respect to the residual momentum is the physical classical velocity

$$\left.\frac{\partial E^{(\tilde{v})}}{\partial p_i}\right|_{\vec{p}=0} = \tilde{v}^{(phys)} \equiv v_i^{(phys)}/v_0^{(phys)} \qquad (5)$$

## 3. EXPANSION IN POWERS OF $v$

The discretization of the HQET reduced Dirac equation we use to put the limiting theory on the lattice is

$$v_0 [U(x,x+\hat{t}) S(x+\hat{t},y) - S(x,y)] + \sum_{\mu=1}^{3} \frac{-iv_\mu}{2} [U(x,x+\hat{\mu}) S(x+\hat{\mu},y) - U(x,x-\hat{\mu}) S(x-\hat{\mu},y)] \qquad (6)$$

$$= \delta(x,y)$$

The use of an asymmetric forward time difference facilitates implementing the requirement that heavy quarks propagate only forward in time. This descretization has the virtue that the heavy quark propagator is obtained by forward recursion. We have described the properties of this implementation of the HQET elsewhere.[3]

The bare classical velocity plays the role of a transverse hopping constant. There is one factor of $\tilde{v}_i$ for each step in the i$^{th}$ direction. After after n time steps the propagator is an $(n-1)^{st}$ order polynomial in the bare classical velocity components. We exploit this structure by developing all quantities that depend on the classical velocity in a power series in the bare components. For example, the meson propagator is given as

$$\Delta^{(v)}(t,\vec{p}) = \sum_{m_1,m_2,m_3} \tilde{v}_1^{m_1} \tilde{v}_2^{m_2} \tilde{v}_3^{m_3} \Delta(t,\vec{p},\vec{m}) \qquad (7)$$

The computation of the coefficients in this polynomial is highly efficient. The indices $m_i$ give the maximum lattice displacement of the heavy quark propagator in the i$^{th}$ direction.

## 4. THE OPTIMIZED MESON FIELD

While in principle any gauge invariant operator containing the reduced heavy quark propagator could be used to extract the physical calssical velocity, in actual simulations it is crucial that the propagator reach its asymptotic form in as few time steps as possible. We therefore choose a composite field which in the Coulomb gauge has the form[4]

$$\Psi^{(v)}(x) = \sum_y \psi(y) q(x) h^{(v)}(x+y) \qquad (8)$$

and determine the weighting function by the requirement that the propagator of this field be maximal on the n$^{th}$ time slice.

The space-time propagator of this composite field is

$$\Delta^{(v)}_{[\psi]}(x,x') = \sum_{y,y'} \psi^*(y) K^{(v)}(x,y;x',y') \psi(y') \qquad (9)$$

where

$$K^{(v)}(x,y;x',y') = \left\langle s(x,x') \tilde{S}^{(v)}(x+y,x'+y') \right\rangle \qquad (10)$$

The average is over the ensemble of lattices. It is amusing to note that while we could "center" the meson field at the heavy quark, the light quark, or anywhere in between, it is handiest to center it at the light quark. The requirement that the propagator be maximal on a given time slice implies that $\psi$ is that eigenvector of $K^{(v)}$ with the largest eigenvalue. The solution to this eigenvalue problem is greatly facilitated by first solving the static, $\vec{p} = 0$, $\tilde{v} = 0$ problem, and then sucessively calculating the coefficients of higher powers of $\tilde{v}_i$.

## 5. SIMULATION OF $v^{(phys)}$

We have computed the magnitude of the classical velocity renormalization by simulating Eq. (9) on a set of $24^3 \times 48$ lattices with $\beta = 6.1$ and $\kappa = .154$ generously provided by the Fermilab Collaboration.[5]

By expanding the logarithmic derivative of the optimized heavy meson residual propagator in powers of the bare classical velocity, we find the combination of terms which should grow linearly with lattice time, with slope equal to the negative of the coefficient of the first power of the bare classical velocity in the physical classical velocity. Taking the finite lattice approximation to the momentum derivative, the appropriate expression is

$$\frac{\Delta(t, p_i = \hat{p}, m_i = 1) - \Delta(t, p_i = -\hat{p}, m_i = 1)}{2\hat{p}\, \Delta(t, \vec{p} = 0, \vec{m} = 0)} \quad (11)$$

where $\hat{p}$ is the minimum momentum on the finite lattice. Figure 1 is the graph of this expression, with its slope, as fitted either over the full 6 time slice range of apparent linearity, or only over the 4 interior slices.

It is interesting to compare the result of this simulation with the shift in the classical velocity evaluated to 1 loop in lattice perturbation theory.[6] For the forward time-symmetric space discretization we have used

$$\delta \tilde{v}_i = \left\{ \begin{array}{ll} -.104 \pm .024 & (5\text{ -- }10\ fit) \\ -.121 \pm .045 & (6\text{ -- }9\ fit) \\ -.2335 & (one\ loop) \end{array} \right\} \tilde{v}_i^{(bare)}$$

One-loop lattice perturbation theory grossly overestimates the renormalization of the classical velocity.

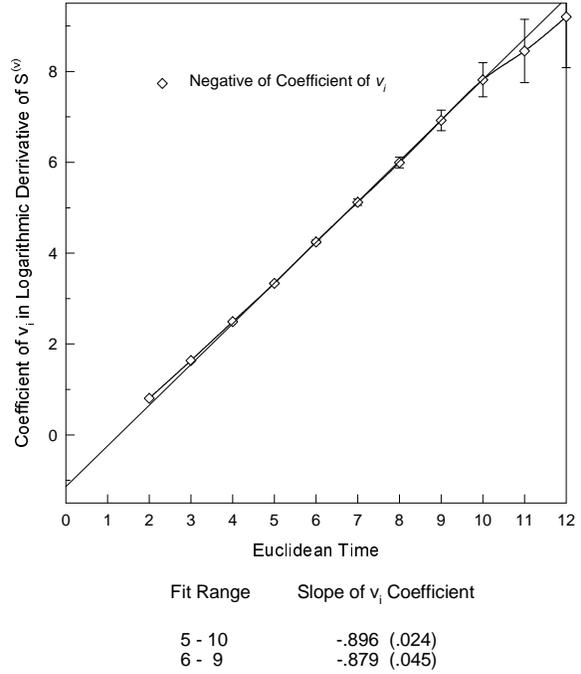

Figure 1 — The expression Eq. (11), and its slope which gives the leading physical classical velocity

| Fit Range | Slope of $v_i$ Coefficient |
|---|---|
| 5 - 10 | -.896 (.024) |
| 6 - 9 | -.879 (.045) |

## ACKNOWLEDGEMENTS

The authors thank the Fermilab ACP-MAPS collaboration for making available to us the ensemble of lattices used in the simulations described here.